\documentclass[aps, twocolumn, prl, longbibliography]{revtex4-1}
\usepackage{amsfonts}
\usepackage{amsmath}
\usepackage{amssymb}
\usepackage{mathptmx} 
\usepackage{graphicx}
\usepackage{bm}
\usepackage{xcolor}
\usepackage[colorlinks=true,linkcolor=blue,anchorcolor=red,citecolor=blue,urlcolor=blue]{hyperref}
\usepackage{ulem}
\usepackage{natbib}
\usepackage{comment}
\begin{document}
\title{Andreev Reflection in the Quantum Hall Regime at an Al/InAs Junction on a Cleaved Edge}
\author{Takafumi Akiho}
\email{takafumi.akiho@ntt.com}
\author{Hiroshi Irie}
\author{Yusuke Nakazawa}
\author{Satoshi Sasaki}
\author{Norio Kumada}
\author{Koji Muraki}
\affiliation{NTT Basic Research Laboratories, NTT Corporation, 3-1 Morinosato-Wakamiya, Atsugi 243-0198, Japan}
\date{\today}

\begin{abstract}
We have fabricated a superconductor/semiconductor (S/Sm) junction composed of Al and InAs using cleaved edge overgrowth. By exploiting the unique geometry with a thin Al/Pt/Al trilayer formed on the side surface of an in-situ cleaved heterostructure wafer containing an InAs quantum well, we achieve a superconducting critical field of $\sim 5$~T, allowing superconductivity and quantum Hall (QH) effects to coexist down to Landau-level filling factor $\nu = 3$. Andreev reflection at zero magnetic field shows a conductance enhancement that is limited solely by the Fermi velocity mismatch, demonstrating a virtually barrier-free, high-quality S/Sm junction. Bias spectroscopy in the QH regime reveals the opening of a superconducting gap, with the reduced downstream resistance demonstrating that the electron-hole Andreev conversion probability consistently exceeds 50\%. Our results, obtained in a new experimental regime characterized by a clean edge-contacted junction with a superconducting electrode narrower than the coherence length, open new avenues for both theoretical and experimental studies of the interplay between superconductivity and QH effects and the engineering of exotic quasiparticles.
\end{abstract}

\maketitle
There have been ongoing efforts to realize topological superconductivity, which is expected to support exotic quasiparticles that obey non-Abelian statistics, a crucial building block for fault-tolerant quantum computation~\cite{FuPRL2008, OregPRL2010, RomanPRL2010, Leijnse2012, Alicea2012,Flensberg2021}.
Among the various proposals to engineer topological superconductivity by exploiting the proximity coupling with $s$-wave superconductors, those based on quantum Hall (QH) systems~\cite{Mong2014, Clarke2013, Clarke2014} are unique in that they hold the promise of otherwise unavailable universal quantum computation using parafermions predicted in the fractional QH regime.
This has led to a renewed interest and experimental efforts~\cite{Wan2015, Amet2016, Lee2017, Kozuka2018, Zhi2019, Guiducci2019, Zhao2020, Wang2021, Gul2022, Hatefipour2022, Zhao2023, Vignaud2023, Zhao2024, Hatefipour2024} in superconductor/semiconductor (S/Sm) junctions in the QH regime, despite the experimental challenges posed by the requirements of junction transparency and the coexistence of high magnetic fields and superconductivity, which are generally incompatible with each other.

Superconducting correlation is induced in QH edge channels by Andreev processes, where electrons entering the junction exit as holes, transmitting Cooper pairs into the superconductor.
Experiments on NbN/graphene junctions in the integer~\cite{Lee2017} and fractional~\cite{Gul2022} QH regimes have demonstrated crossed Andreev reflection, in which electron-hole conversion takes place between counterpropagating edge channels through a narrow strip of superconductor in between.
There, the negative resistance in the downstream of the junction ($R_\text{d}$) provided evidence of electron-hole conversion.
In contrast, for a 600-nm-wide MoRe/graphene junction along a single edge, $R_\text{d}$ was observed to oscillate or fluctuate between positive and negative as the magnetic field or gate voltage was swept~\cite{Zhao2020, Zhao2023}.
It was then argued that these oscillations arise from the interference of chiral Andreev edge states (CAESs), hybrid eigenmodes of electron- and hole-like edge states formed along the proximitized region~\cite{Hoppe2000,Giazotto2005}.
However, subsequent theoretical investigations have raised interfacial disorder~\cite{Manesco2022}, vortices~\cite{Tang2022}, disorder~\cite{Kurilovich2023}, and supercurrent~\cite{Michelsen2023} in the superconductor, and the geometry of the junction~\cite{David2023} as necessary ingredients to explain the experimental observations.
On the other hand, an experiment for a 150-$\mu$m-wide NbTiN/InAs junction found negative $R_\text{d}$ without oscillations~\cite{Hatefipour2022}.
These clearly show the need for further experiments with a controlled setup to understand even the most fundamental process of Andreev conversion.

Common to the reported superconducting junctions in the QH regime~\cite{Takayanagi1998, Eroms2005, Rickhaus2012, Wan2015, Amet2016, Lee2017, Kozuka2018, Zhi2019, Zhao2020, Gul2022, Hatefipour2022, Zhao2023, Vignaud2023} is that they use type-II superconductors to sustain superconductivity at high fields and etching to fabricate the junction, both of which make it challenging to form a pristine and uniform junction free of interfacial disorder or in-gap states.
Nonuniform interfaces would result in an inhomogeneous induced gap, leading to a soft gap~\cite{Takei2013}, where in-gap states act as a source of unwanted quasiparticle tunneling into the superconductor.
On the other hand, successful formation of pristine S/Sm junctions by in-situ epitaxial growth of a superconductor on the surface of a semiconductor has been reported~\cite{Krogstrup2015, Shabani2016, Bjergfelt2019, Kanne2021, Khan2023}.
In particular, epitaxial junctions of Al and InAs nanowires~\cite{Chang2015} or two-dimensional electron systems (2DESs)~\cite{Kjaergaard2016} have been shown to have a hard superconducting gap free of in-gap states.
Although Al is a type-I superconductor, in thin films it can maintain superconductivity up to a few tesla when the field is in the plane of the film, a property that has been exploited for hybrid nanowires.
Furthermore, by adding heavy elements with strong spin-orbit coupling such as Pt, the in-plane critical field can be increased to a several tesla~\cite{Meservey1994,Mazur2022}.
However, despite their superior properties, epitaxial Al/InAs junctions could not be used for QH systems, because the field must be applied perpendicular to the 2DEG, i.e., to the superconducting film.

Here, we overcome this limitation by using a novel geometry in which the superconducting thin film is grown on the side surface of a semiconductor wafer, and realize an Al/InAs epitaxial junction that operates in the QH regime.
Using the cleaved edge overgrowth (CEO) technique~\cite{Pfeiffer1993},
we deposited an Al/Pt/Al trilayer on an in-situ cleaved edge surface of a wafer containing an InAs quantum well (QW).
Andreev reflection at zero field indicates that the conductance enhancement is limited solely by the Fermi velocity mismatch, demonstrating a high-quality junction that is virtually free of an interface barrier.
At high fields, we observe that $R_\text{d}$ is reduced near zero bias in the plateau regions for filling factors from 10 to 3, indicating that the probability of electron-hole conversion consistently exceeds 50\%, without oscillatory behavior.
The geometry of our experiment is unique in that 1) the field direction, being parallel to the plane of the thin superconductor, eliminates vortices, and 2) the 2DES is directly edge-contacted by the superconductor, with no spatial overlap along the field direction.
The observation of Andreev conversion with a clear bias dependence in the unique geometry provides new insights into the underlying processes.
In addition, our approach to fabricating etch-free, high-quality superconducting junctions will open a new avenue for the study of engineered topological superconductivity.

Our CEO device was fabricated from a heterostructure wafer grown on a GaSb(001) substrate. (See Supporting Information (SI) for the layer structure.)
The wafer, cut into a $14$ mm $\times 15$ mm piece, was thinned to $300~\mu$m by polishing to facilitate cleaving and thus obtain a flat cleaved edge surface.
The wafer was mounted at 90$^\circ$ on a Mo sample holder and was cleaved in UHV ($< 5.0 \times 10^{-10}$~mbar) using a wobble stick.
The wafer was then transferred under UHV to the deposition chamber, where a superconducting electrode was deposited on the cleaved edge surface, with the sample holder cooled by direct contact with a liquid nitrogen cold plate, a crucial step for obtaining epitaxial Al layers~\cite{Krogstrup2015}.
We used photolithography to process the sample into a gate-defined Hall bar with normal Ti/Au electrodes and an S/Sm junction at the cleaved edge.
(See SI for fabrication details.)
Measurements were conducted in a $^{3}$He cryostat equipped with a superconducting vector magnet.
The 2DES density and mobility are $n = 3.0 \times 10^{15}$m$^{-2} $ and $\mu = 63~\text{m}^{2}$/Vs, respectively.

\begin{figure}
  \includegraphics[scale=1.0]{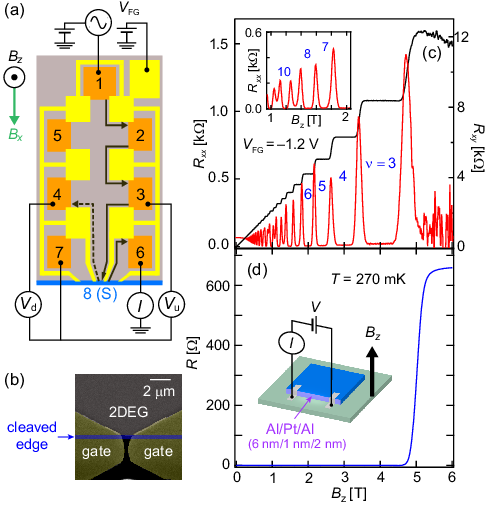}
  \caption{(a) Schematic representation of the device and the measurement setup. \#1 through \#7 and \#8 represent normal and superconducting electrodes, respectively. The black solid and dashed lines with arrows indicate the direction of the edge channels under a perpendicular magnetic field $B_{z}$. (b) SEM image of the S/Sm interface at the cleaved edge of a dummy sample. (c) Magnetic field dependence of $R_{xx}$ (red) and $R_{xy}$ (black) measured using normal contacts. The inset shows a magnified view of $R_{xx}$ at low fields. (d) Magnetic field dependence of the resistance of a control Al/Pt/Al sample. The inset shows a schematic of the measurement setup. The magnetic field is applied parallel to the Al/Pt/Al film.}
  \label{figure1}
\end{figure}

Figure~\ref{figure1}(a) shows a schematic of our CEO device.
The Hall bar is defined electrostatically by applying a negative voltage $V_\text{FG}$ ($\lesssim -1.2$~V) to the front gate and depleting the 2DES beneath~\footnote{This prevents non-topological edge conduction in the QH regime caused by electron accumulation at the edge, which can occur in etch-defined InAs channels~\cite{vanWees1995, Akiho2019, Komatsu2022}. Additionally, it allows the junction ends to be defined electrostatically, avoiding contamination or damage from direct processing.}.
The front gate has an opening above the cleaved edge, defining a narrow junction of width $L$, nominally 3-$\mu$m wide, between the superconductor (\#8) and the 2DES.
The scanning electron microscope (SEM) image taken near the narrow junction of a dummy sample confirms the formation of a split gate [Fig.\,\ref{figure1}(b)].
The superconductor (\#8) was electrically accessed through the wide S/Sm junctions on either side, using the 2DES regions isolated from the Hall bar by the front gate as leads and the Ti/Au contacts \#6 and \#7 as current and voltage probes, respectively~\footnote{This design was employed to access the superconductor while avoiding unintended breaks at steps on the cleaved edge and shorting to the conducting GaSb substrate.}.
We applied AC and DC bias to the source electrode \#1 and measured the AC current ($I$) from the drain electrode \#6.

Figure~\ref{figure1}(c) presents the longitudinal ($R_{xx}$) and Hall ($R_{xy}$) resistances of the device, measured in a perpendicular field $B_{z}$ using normal contacts, showing well-developed QH effects at Landau-level filling factor $\nu$ ($= nh/eB_{z}$) of 3 to 10, where $h$ is Planck's constant and $e$ is the elementary charge.
Figure~\ref{figure1}(d) shows the superconducting property of the Al/Pt/Al film of a control sample fabricated on the cleaved edge of a semi-insulating GaAs(100) substrate.
The two-terminal resistance $R$ of an Al/Pt/Al trilayer of thicknesses 6, 1, and 2~nm from bottom to top is plotted as a function of the field $B_{z}$ applied perpendicular to the substrate (i.e., parallel to the superconducting thin film).
The critical in-plane field $B_\text{c}^{\parallel}$ is $\sim5$~T for the 6/1/2-nm trilayer, which decreased as the bottom Al thickness increased.
Based on this result, we employed the 6/1/2-nm trilayer structure for the S/Sm junction.
The results in Figs.\,\ref{figure1}(c) and (d) together confirm that, below $B_{z} = 5$~T, superconductivity and QH effects can coexist in our device employing an Al superconductor.

\begin{figure}
  \includegraphics[scale=1.0]{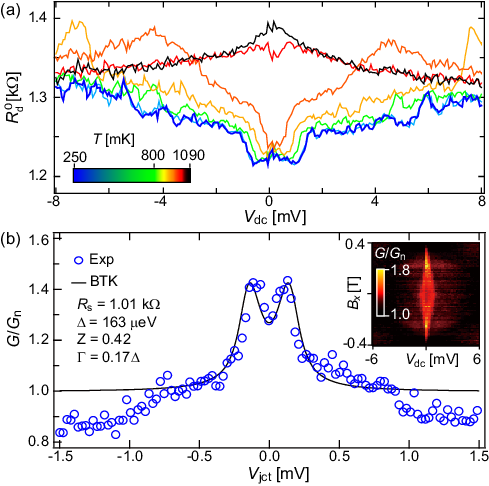}
  \caption{(a) DC bias dependence of the differential resistance at zero magnetic field and at different temperatures. (b) Bias dependence of the differential conductance, normalized to the value outside the superconducting gap. Note that the horizontal axis represents the voltage across the junction $V_\text{jct}$ [$\equiv (\Delta/e)V_\text{dc}/V_\Delta$] (see main text for details). The solid line represents a fit using the BTK model. (Inset) Normalized differential conductance as a function of $V_\text{dc}$ and $B_{x}$, with the magnetic field $B_{x}$ applied perpendicular to the superconducting film.}
  \label{figure2}
\end{figure}

Now we characterize the S/Sm junctions by measuring the differential resistance at $B_{z} = 0~$T using the setup shown in Fig.\,\ref{figure1}(a).
Here, we focus on the results for $R^{0}_{\text{d}} = \text{d}V_\text{d}/\text{d}I$ obtained with contacts \#4 and \#7, while those for $R^{0}_{\text{u}} = \text{d}V_\text{u}/\text{d}I$ obtained with contacts \#3 and \#7 are provided in SI.
Figures~\ref{figure2}(a) show the DC bias ($V_\text{dc}$) dependence of $R^{0}_{\text{d}}$ at different temperatures~\footnote{Throughout this paper, $V_\text{dc}$ is the measured DC component of $V_\text{u}$, the volatage between contacts \#3 and \#7.}.
A pronounced zero-bias minimum is observed at low temperatures, demonstrating Andreev reflection, i.e., reflection of an incident electron from the 2DES as a hole, thereby enhancing the conductance~\cite{Andreev1964}.
From the temperature at which the zero-bias minimum evolves into a broad maximum, the critical temperature $T_\text{c}$ is determined to be $1070$~mK, with the corresponding BCS gap $\Delta = 1.764k_\text{B}T_\text{c}$ of $163~\mu$eV, where $k_\text{B}$ is the Boltzmann constant.

To characterize the S/Sm junction more quantitatively, we evaluate the Andreev reflection probability.
To this end, we derived the differential conductance of the junction, $G = (R^{0}_\text{d} - R_\text{s})^{-1}$, by subtracting the series resistance $R_\text{s}$ from $R^{0}_\text{d}$.
We estimated $R_\text{s} = 1.01~$k$\Omega$ by noting that the voltage drop across the S/Sm junction ($V_\text{jct}$) equals $\Delta/e$ at the DC bias ($V_\Delta$) for which the Fermi level in the normal region coincides with the superconducting gap edge at base temperature.
Figure~\ref{figure2}(b) plots the normalized differential conductance $G/G_\text{n}$ as a function of $V_\text{jct}$ [$\equiv (\Delta/e)V_\text{dc}/V_\Delta$], where $G_\text{n}$ is the differential conductance at a finite bias outside the superconducting gap.
The solid line represents a fit based on the modified Blonder-Tinkham-Klapwijk (BTK) model described by~\cite{BlonderPRB1982, Dynes1987, Naidyuk1996, Lee2017},
\begin{equation}
G/G_\text{n} = C\int_{-\infty}^{\infty} \left\lbrack f(E-eV) - f(E) \right\rbrack (1 + A - B) \text{d}E.
\end{equation}
Here, $C$ is a constant, $E$ is the energy, $f$ is the Fermi distribution function, and $A = A(E, \Delta, Z, \Gamma)$ and $B = B(E, \Delta, Z, \Gamma)$ are the Andreev and normal reflection probabilities, respectively
\footnote{
$A = a^{\ast} a$ and $B = b^{\ast} b$, where $a = u_{0} v_{0} /\gamma$, $b = -(u_{0}^{2} -  v_{0}^{2}) /\gamma$, and $\gamma = u_{0}^{2} +  Z^{2}(u_{0}^{2} - v_{0}^{2})$ with $u_{0}^{2} = \frac{1}{2} \Big[ 1 + \frac{\sqrt{(E + i \Gamma)^{2} - \Delta^{2}}}{E + i \Gamma} \Big]$ and $v_{0}^{2} = \frac{1}{2} \Big[ 1 - \frac{\sqrt{(E + i \Gamma)^{2} - \Delta^{2}}}{E + i \Gamma}\Big]$
}.
$Z$ is a dimensionless parameter characterizing the effective barrier at the S/Sm interface, and $\Gamma$, which enters $A$ and $B$ as the imaginary part of the energy, describes the lifetime broadening.

As shown in Fig.\,\ref{figure2}(b), $Z = 0.42$ and $\Gamma = 0.17\Delta$ reproduce the measured bias dependence, including the conductance enhancement factor $G/G_\text{n}$ of $1.28$ at zero bias.
Both finite $Z$ and $\Gamma$ reduce the Andreev reflection probability, resulting in a conductance enhancement factor less than 2, the value expected for an ideal junction.
Note, however, that a mismatch in the Fermi velocity at the junction leads to a finite normal reflection probability even in the absence of a barrier.
The resulting $I$-$V$ characteristics can be described by replacing $Z$ with an effective value $Z_\text{eff}=\sqrt{Z^{2}+(1-r)^{2}/4r}$, where $r$ is the Fermi velocity ratio~\cite{Blonder1983}.
In fact, for our InAs 2DES with $v_\text{F} = 5 \times 10^{5}$~m/s, the Fermi velocity mismatch alone is sufficient to make $Z_\text{eff} \geq 0.42$, assuming bulk Al~\cite{Wegehaupt1977}.
Including the two-dimensionality of the 2DES, which is not considered in the BTK model, would further increase $Z_\text{eff}$~\cite{Mortensen1999}.
These observations suggest that our junction is virtually barrier-free.
Interestingly, the obtained $\Gamma = 0.17\Delta$ corresponds to a quasiparticle lifetime $\tau_\text{qp}$ ($= h/2 \pi \Gamma$) of $24$~ps.
This is more than 10 times longer than that for the NbN/graphene device~\cite{Lee2017}, indicating that a high-quality superconducting electrode has been obtained by CEO.

We also measured Andreev reflection with the magnetic field $B_{x}$ applied perpendicular to the superconducting thin film [inset of Fig.\,\ref{figure2}(b)].
From the critical field of $B_\text{c}^{\perp} = 0.39$~T, the effective coherence length is estimated to be $\xi = 29~$nm.
While this value is much shorter than the typical coherence length for Al ($> 1~\mu$m), possibly due to the spin-orbit scattering induced by Pt, it is important to note that the thickness of our superconducting film (10~nm) is even smaller.

\begin{figure}[ptb]
\includegraphics[scale=1.0]{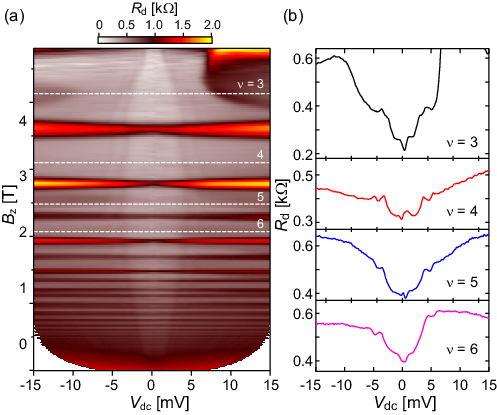}
  \caption{(a) Color scale plot of $R_\text{d}$ as a function of $V_\text{dc}$ and $B_{z}$. (b) Line cuts of the data in (a) at $\nu = 3$, 4, 5, and 6.}
  \label{figure3}
\end{figure}

We next study the Andreev reflection in the QH regime by applying a magnetic field $B_{z}$ perpendicular to the 2DES (i.e., parallel to the superconducting thin film).
Figure~\ref{figure3}(a) plots $R_\text{d}$, the resistance measured on the edge channels downstream of the S/Sm junction, as a function of $V_\text{dc}$ and $B_{z}$.
The plot reveals a region of reduced $R_\text{d}$ around $V_\text{dc} = 0$~V, which narrows above 4~T but remains open up to 4.8~T.
We associate this with the superconducting gap~\footnote{The width of the $V_\text{dc}$ range corresponding to the superconducting gap becomes narrower at low fields due to the increasing bulk conductivity near the S/Sm junction.}.
Figure~\ref{figure3}(b) displays horizontal line cuts of Fig.\,\ref{figure3}(a) at several integer values of $\nu$, showing reduced $R_\text{d}$ inside the superconducting gap.
We emphasize that such a clear bias dependence, indicating a superconducting gap in the QH regime, has not been previously reported for junctions along a single edge~\footnote{The only reports are for the setups of crossed Andreev reflection~\cite{Lee2017, Gul2022}.}.
We first notice that although $R_\text{d}$ is clearly reduced around zero bias, it remains positive.
This contrasts with the sign-oscillating $R_\text{d}$ reported for MoRe/graphene junctions~\cite{Zhao2020, Zhao2023}, where negative $R_\text{d}$ indicates electron-hole conversion.
We note that in our data $R_\text{d}$ is finite outside the superconducting gap.
This indicates that the coupling between the edge channels and the Al electrode is imperfect; that is, even in the absence of superconductivity, some fraction of electrons in the upstream channels are transferred to the downstream channels.
As we show below, this effect can be incorporated into a Landauer-B\"{u}ttiker-type model, from which the probability $P_\text{h}$ of electron-hole conversion by the Andreev process can be derived.

\begin{figure}[ptb]
  \includegraphics[scale=1.0]{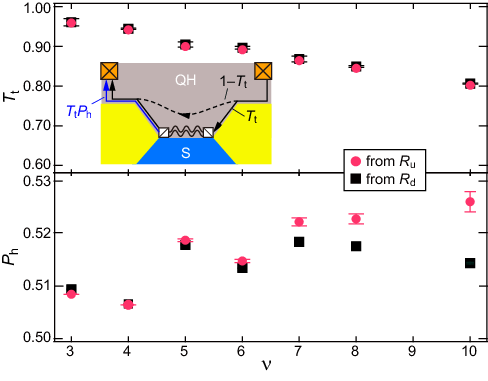}
  \caption{$\nu$ dependence of $T_\text{t}$ (upper panel) and $P_\text{h}$ (lower panel). The plots show results derived separately from $R_\text{d}$ and $R_\text{u}$ using Eqs.~(3) and (4), respectively. The inset shows a schematic of the processes involved in our Andreev reflection measurement in the QH regime. The black solid lines indicate the flow of electrons entering and exiting the S/QH junction. The dashed line schematically shows the path of electrons traveling downstream without impinging on the junction. The wavy lines depict CAESs, while the blue line illustrates the flow of holes converted from electrons through the CAESs.}
  \label{figure4}
\end{figure}

We construct a Landauer-B\"{u}ttiker-type model appropriate for our device layout involving a superconducting electrode.
The current-voltage relation can be expressed as $\mathbf{I} = \nu G_{0} \mathbf{M} \mathbf{V}$, where $G_{0} = e^{2}/h$ is the conductance quantum, and $\mathbf{I} = (..., I_{i}, ...)^\mathrm{T}$ and $\mathbf{V} = (..., V_{i}, ...)^\mathrm{T}$ with $I_{i}$ ($V_{i}$) the current (voltage) of the $i$th contact ($i = 1$, 3, 4, 6, 7, 8).
(Contacts \#2 and \#5, which are not used in the $R_\text{d(u)}$ measurements, are excluded.)
$\mathbf{M}$ is the normalized conductance matrix.
The effect of imperfect edge coupling is accounted for by introducing a parameter $T_\text{t}$.
As shown schematically in the inset of Fig.\,\ref{figure4}, a portion of the electrons in the incoming $\nu$ edge channels are reflected before reaching the superconducting electrode \#8 with probability $1-T_\text{t}$ and fed into the outgoing channels.
The definition of $P_\text{h}$ is such that the injection of an electron into the junction results in the emission of, on average, $P_\text{h}$ holes into the outgoing channels.
That is, $P_\text{h} = 1$ and $0$ correspond to 100\% electron-hole conversion (i.e., perfect Cooper-pair transmission) and perfect normal reflection (i.e., no current through the junction), respectively.
$P_\text{h} = 0.5$ means that on average these two events occur with equal probability.
Thus, the matrix $\mathbf{M}$ is given as
\begin{equation}
\mathbf{M} =
\begin{pmatrix}
1 & 0 & -1 & 0 & 0 & 0 \\
-1 & 1 & 0 & 0 & 0 & 0 \\
0 & \eta-1 & 1 & 0 & 0 & -\eta \\
0 & 0 & 0 & \eta^\prime & 0 & -\eta^\prime \\
0 & 0 & 0 & 0 & \eta^\prime & -\eta^\prime \\
0 & -\eta & 0 & -\eta^\prime & -\eta^\prime & \eta+2\eta^\prime \\
\end{pmatrix},
\end{equation}
where $\eta \equiv 2T_\text{t}P_\text{h}$ and $\eta^\prime$ is the counterpart of $\eta$ for the wide junctions connected to contacts \#6 and \#7.
By solving $\mathbf{I} = \nu G_{0} \mathbf{M} \mathbf{V}$ for $\mathbf{V}$ with $\mathbf{I}=(-I,0,0,I,0,0)^\mathrm{T}$, we obtain~\footnote{The consistency of the model can be checked by seeing that $T_\text{t} = 1$ and $P_\text{h} = 0.5$ give $R_\text{d} = 0$ and $R_\text{u} = 1/\nu G_{0}$ as expected for the conventional QH effects with normal contacts, while $T_\text{t} = 1$ and $P_\text{h} = 1$ give $R_\text{d} = -1/2 \nu G_{0}$ and $R_\text{u} = 1/2 \nu G_{0}$ as expected for perfect Andreev reflection.}
\begin{align}
	R_\text{d} &\equiv \frac{V_{4}-V_{7}}{I}= \left( \frac{1}{2T_\text{t}P_\text{h}} - 1 \right) \frac{1}{\nu G_{0}},\\
	R_\text{u} &\equiv \frac{V_{3}-V_{7}}{I} = \frac{1}{2T_\text{t}P_\text{h}} \frac{1}{\nu G_{0}}.
\end{align}
Note that for $T_\text{t}<1$, $R_\text{d}$ becomes negative only when $P_\text{h} > 1/2T_\text{t}$, even if $P_\text{h} > 0.5$.

We deduced $T_\text{t}$ at each integer $\nu$ from the measured $R_\text{d}$ and $R_\text{u}$ outside the superconducting gap ($V_\text{dc} = -10$~mV) by setting $P_\text{h}= 0.5$ in Eqs.\,(3) and (4).
As shown in the upper panel of Fig.\,\ref{figure4}, $R_\text{d}$ and $R_\text{u}$ give consistent results, both showing that $T_\text{t}$ gradually increases from 0.8 to 0.96 as $\nu$ decreases from 10 to 3.
We used these $T_\text{t}$ values to derive $P_\text{h}$ from $R_\text{d}$ and $R_\text{u}$ at $V_\text{dc} = 0$~mV.
As shown in the lower panel of Fig.\,\ref{figure4}, $R_\text{d}$ and $R_\text{u}$ give similar results, showing that $P_\text{h}$ consistently exceeds $0.5$.

Two main aspects of our results need to be discussed: the absence of oscillatory behavior in $R_\text{d}$ and the preferential reflection of holes, indicated by $P_\text{h}>0.5$.
Theories predict that, for an ideal junction without disorder, $P_\text{h}$ should oscillate as a function of junction width $L$~\cite{Kurilovich2023,Arrachea2024,Van_Ostaay2011} and bias~\cite{Gamayun2017,Tang2022}, due to the interference between CAESs, reflecting their momentum difference.
On the other hand, it is also argued that the oscillations will only be visible at specific magnetic fields per Landau level; away from these fields, the oscillation amplitude will rapidly decay, causing $P_\text{h}$ to average out to 0.5~\cite{Kurilovich2023}.
The absence of such field dependence in the reported experiment~\cite{Zhao2020} is a key aspect that triggered the subsequent theoretical investigations~\cite{Manesco2022,Tang2022,Kurilovich2023,Michelsen2023,David2023}.
In our data, $R_\text{d}$ (and hence $P_\text{h}$) remains nearly constant within each QH plateau, maintaining $P_\text{h} >0.5$, without  oscillations or sharp field dependence.
One could speculate on the reasons for the absence of oscillations, such as the involvement of multiple edge channels or factors like temperature, momentum difference, or junction width being inadequate for resolving the interference.
However, in either case, the question arises as to why $P_\text{h}$ does not average out to $0.5$.
The absence of sharp field dependence also remains unexplained.

Turning to the result in Fig.~\ref{figure4}(b), $P_\text{h}$ gradually increases with increasing $\nu$ (i.e., decreasing $B_{z}$).
No clear difference in $P_\text{h}$ is observed between the odd and even $\nu$, where the incoming edge channels are partially spin-polarized and unpolarized, respectively.
This suggests the presence of a spin-flip mechanism, likely due to the spin-orbit interaction in the Pt layer.

Another notable feature in our data is the absence of discontinuous jumps in $R_\text{d}$ during magnetic field sweeps, which were reported for MoRe/graphene junctions and attributed to a phase shift in $\Delta$ resulting from vortices hopping in and out of the superconductor~\cite{Zhao2020, Zhao2023}.
In our setup, the superconductor is thinner than the coherence length, and we utilize a two-axis vector magnet to align the magnetic field parallel to the film. This minimizes the possibility of vortices playing any role, consistent with the absence of discontinuous jumps.
In an attempt to account for the $P_\text{h} > 0.5$ observed in their 150-$\mu$m-wide NbTiN/InAs junction, the authors of Ref.~\cite{Hatefipour2022} speculated that electrons may have a higher probability than holes to tunnel into the superconductor.
Although the origin of such asymmetry, if present, is unclear, the absence of vortices, which promote electron/hole tunneling into the superconductor, makes this scenario less likely in our CEO device.
While the effects of vortices have been studied from different theoretical perspectives~\cite{Kurilovich2023, Tang2022}, the irrelevance of vortices in our CEO device is expected to help elucidate the mechanism of Andreev conversion in the QH regime.

The narrowness of the superconducting electrode can affect the Andreev processes in several additional ways. These include the guiding center of the electron and hole orbits comprising the CAESs~\cite{Hoppe2000, Giazotto2005, Kurilovich2023}, confinement-induced quantization in the superconductor~\cite{Nguyen2019, Stanescu2022}, and the spatial distribution of the screening supercurrent.
Furthermore, the much smaller $\Delta$ of Al compared to those in previous studies would affect the velocity of CAESs and their momentum difference.
These factors place our experiment in a unique regime, distinct from previous ones.
On the other hand, it is argued that the scattering properties at the endpoints of the junction play a key role in determining the electron-hole conversion probability, which, however, cannot be adequately accounted for by one-dimensional models~\cite{David2023}.
Two-dimensional models that properly incorporate the above effects will be necessary to understand the observed behavior.
Our observation of $P_\text{h} > 0.5$ along with a clear bias dependence in a unique setup will thus stimulate further theoretical and experimental investigations.

In summary, using cleaved edge overgrowth, we have successfully fabricated a clean, virtually barrier-free Al/InAs junction that operates in the QH regime.
Thanks to the high interface quality, we observed the opening of a superconducting gap in bias spectroscopy up to 4.8~T in the QH regime---a behavior previously unseen in junctions along a single edge.
The reduced downstream resistance within the superconducting gap indicates an electron-hole Andreev conversion probability consistently exceeding 50\%.
Our results, obtained in a unique setup where the 2DES is directly edge-contacted by a superconducting electrode narrower than the coherence length, are expected to stimulate further theoretical investigations into the processes involved in Andreev conversion.
Our demonstration of an epitaxial S/QH junction utilizing Al, a significantly cleaner superconductor, opens new avenues for exploring non-Abelian quasiparticles, with substantial implications for both experimental and theoretical advancements in nanoscience.


The authors thank H. Murofushi for device processing, M. Imai for drawing support, H. Kamata for measurement advice, and T. Wakamura for discussions on Andreev reflection data.


%

\end{document}